\documentclass[showpacs,preprint,amsmath,amssymb]{revtex4}

\usepackage{graphicx}
\usepackage{epsfig}
\usepackage{xcolor}
\usepackage{dcolumn}
\usepackage{longtable}
\usepackage{longtable}
\usepackage{tabularx}
\usepackage{array}

\begin{document}

\title{Angular momentum dependence of $\alpha$-decay spectroscopic factors}

% Force line breaks with \\
%\thanks{A footnote to the article title}%
\author{E. V. Mardyban}
\altaffiliation[Also at ]{Korkyt Ata Kyzylorda University, Kyzylorda, Kazakhstan}
\altaffiliation[Also at ]{Dubna University, Dubna, Russia}
\affiliation{Joint Institute for Nuclear Research, Dubna, 141980, Russia}
\author{D. F. Bayramov}
\affiliation{Moscow state university, Moscow, Russia}
\altaffiliation[Also at ]{Joint Institute for Nuclear Research, Dubna, 141980, Russia}

\author{A. Rahmatinejad}
\affiliation{Joint Institute for Nuclear Research, Dubna, 141980, Russia}

\author{A.~K.~Azhibekov}
\altaffiliation[Also at ]{Korkyt Ata Kyzylorda University, Kyzylorda, Kazakhstan}
\altaffiliation[Also at ]{Institute of Nuclear Physics, Almaty, Kazakhstan}
\affiliation{Joint Institute for Nuclear Research, Dubna, 141980, Russia}

\author{K. Mendibayev}
\altaffiliation[Also at ]{Korkyt Ata Kyzylorda University, Kyzylorda, Kazakhstan}
\altaffiliation[Also at]{Institute of Nuclear Physics, Almaty, Kazakhstan}
\affiliation{Joint Institute for Nuclear Research, Dubna, 141980, Russia}
\author{T. M. Shneidman}
%\altaffiliation[Also at ]{Kazan Federal University, Kazan 420008, Russia}
\affiliation{Joint Institute for Nuclear Research, Dubna, 141980, Russia}

\date{\today}% It is always \today, today,
             %  but any date may be explicitly specified

\begin{abstract}
The probabilities to form an $\alpha$-particle on the surface of heavy nuclei (spectroscopic factors) in the states of different angular momenta are
calculated within the dinuclear system (DNS) approach. It is shown that this dependence is determined not only by the energies of the excited states of the daughter nucleus, as is suggested by the Boltzmann distribution, but also by its quadrupole and octupole deformation parameters.
The calculations were performed  for actinide nuclei $^{222-228}$Ra, $^{222-232}$Th, $^{226-236}$U, and $^{234-242}$Pu. The results obtained will be useful in the future studies on the fine structure of alpha-decay.
\end{abstract}

\pacs{}

\maketitle

\section{introduction}

One of the main modes of decay of atomic nuclei in the region of actinides and superheavy nuclei is $\alpha$-decay. A widely used assumption in solving the problem of $\alpha$-decay is that the $\alpha$-particle is formed with some probability on the surface of the parent nucleus. The $\alpha$-particle is confined in a potential well formed by attractive nuclear and repulsive Coulomb potentials. Since the potential barrier has a finite height, there is the possibility of spontaneous decay of the system. The probability of $\alpha$-decay is thus determined by the product of two factors: the spectroscopic factor, which determines the probability of formation of an $\alpha$-particle on the surface of the nucleus and the penetrability of the potential barrier. This approach, based on the pioneering work of G. Gamow \cite{Gamow1928}, turned out to be very successful in describing the main characteristics of $\alpha$-decay. For example, the Geiger-Netall law was established, which connects the half-life with the energy $Q_\alpha$ released in $\alpha$-decay \cite{Kadmensky1985}.
This established relationship effectively explains the regular changes in the probabilities of $\alpha$-decay between the ground states of even-even nuclei. However, for transitions to excited states, sharp changes in the decay rates are observed \cite{Gallagher1957}.

A possible explanation for this behavior is the dependence of both the tunneling probability and the spectroscopic factor on the deformation of the parent nucleus. First, it should be taken into consideration that if the daughter nucleus is deformed, then the potential barrier is not spherically symmetric. As a consequence, the angular momentum of the $\alpha$-particle can change during tunneling \cite{Kuklin2012}. Secondly, the deformation of the daughter nucleus leads to the fact that the $\alpha$-particle is formed on the surface of the nucleus in states with different angular momenta.
The interplay of these two factors determines the resulting fine structure of $\alpha$-decay.

Two primary approaches are used to solve the $\alpha$-decay problem: the semiclassical \cite{Ren2006, Santosh2011, Denisov, Seif2023}  and the microscopic\cite{Ren2010, Ren2011} approaches, with the latter employing the coupled-channel formalism.
 However, so far, the angular momentum dependence of the spectroscopic factor has not been taken into account explicitly. It is either assumed that all angular momenta are populated with equal probabilities as in Ref. \cite{Kuklin2012,Denisov}, or treated empirically using the Boltzmann distribution (BD) \cite{Seif2023,Ren2006}.

In this work, we directly calculate the angular momentum dependence of spectroscopic factors using the dinuclear system model (DNS) \cite{Shneidman2003,Shneidman2015}. The role of quadrupole and octupole deformation of the daughter nucleus is analyzed. The obtained results are compared with the empirical values provided by BD.

\section{Model}

In the actinide mass region, the wave function of a nucleus can be represented as a superposition of the mononuclear configuration
and the alpha-cluster dinuclear system (DNS) $\Psi_\alpha$ \cite{Shneidman2003}:
\begin{eqnarray}
\Psi = \cos\gamma \Psi_m + \sin \gamma \Psi_\alpha,
\label{Psi-Stationary}
\end{eqnarray}
where the weight of $\alpha$-cluster component (spectroscopic factor) is defined as  $S_{\alpha}=\sin^2{\gamma}$. To obtain the spectroscopic factor $S_{\alpha}$ one can solve the Schr\"odinger equation in mass asymmetry coordinate (see, for example, Refs.~\cite{Shneidman2015,Kuklin2012}).  The spectroscopic factor depends on relative distance between the centers of the daughter nucleus and the $\alpha$ -particle.  At distances less than touching, the $\alpha$ -particle dissociates (see Ref.~\cite{Schuck2016}) and the spectroscopic factor decreases to zero. For larger distances, the spectroscopic factor fast approaches unity. At touching, the probability to form an $\alpha$-particle on the surface of the heavy nucleus ($\alpha$-DNS) achieves several percents. As was shown in~\cite{Shneidman2015}, the contribution of this component is crucial to describe the formation of negative parity states in actinides. In the current calculations, however, we are not interested in the spectroscopic factor itself but rather its dependence on the angular momentum. Since the mononucleus wave function does not contribute to the $\alpha$-decay, this dependence originates from the structure of $\alpha$-DNS wave function $\Psi_\alpha$.

\begin{figure}[bth]
\begin{center}
\includegraphics[scale=.5 ]{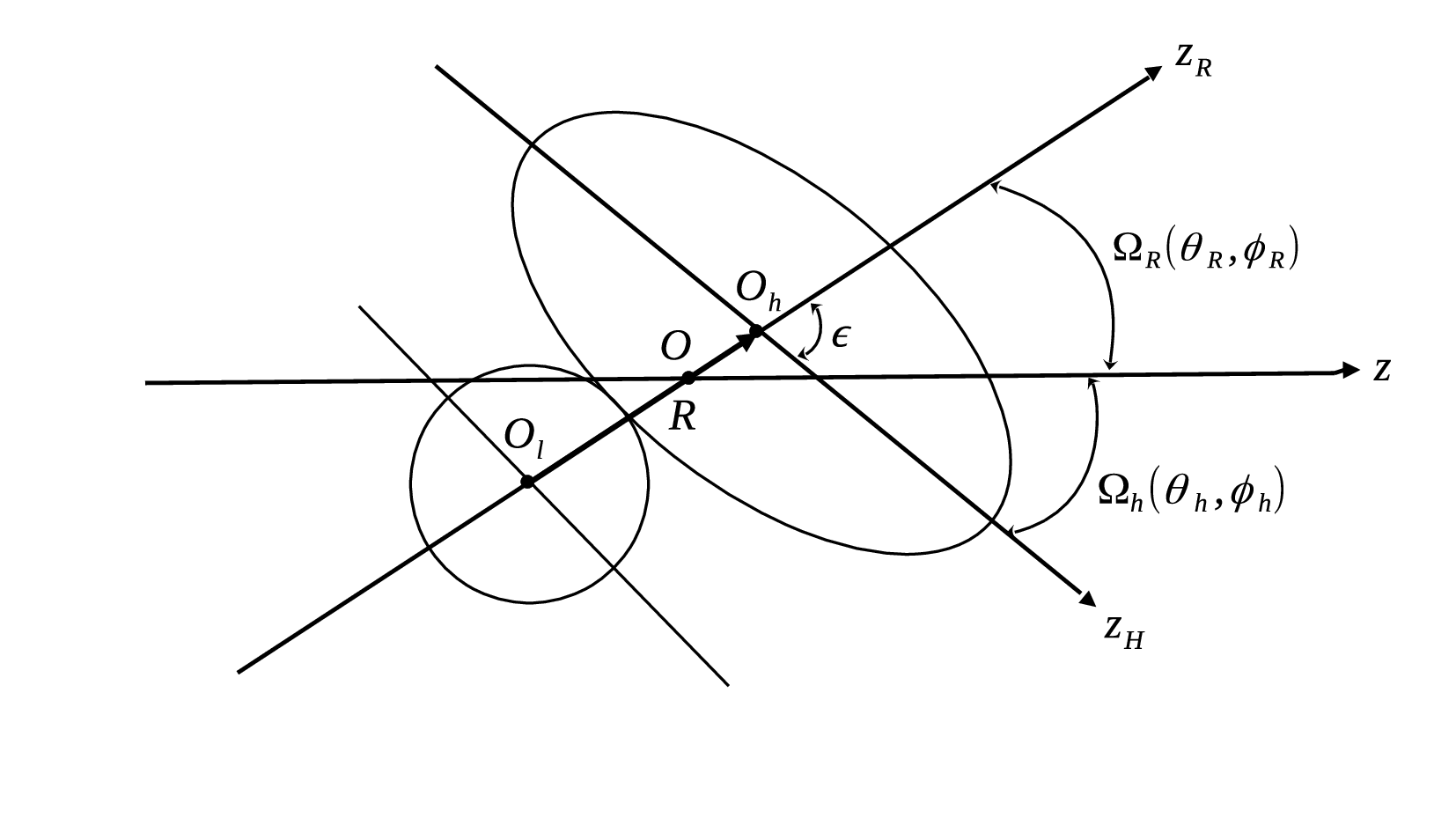}
\end{center}
\vspace{-.5cm} \caption{
Schematic representation of $\alpha$-DNS.
%with indication of the degrees of freedom used.
The orientation of the vector  ${\bf R}$, connecting the centers of two fragment, is defined by
the angles $\Omega_R=(\theta_R, \phi_R)$ with respect to the laboratory system  $Oz$. The orientation of
the intrinsic coordinate system of the quadrupole-deformed fragment $O_h z_h$ with respect to the laboratory
system is described by the angles  $\Omega_h=(\theta_h, \phi_h).$
The angle $\epsilon$  is a plane angle between  ${\bf R}$ and the symmetry axis of the deformed fragment.}
\label{coor}
\end{figure}

In the following, we assume that only the lowest collective excited states contribute to $\Psi_{\alpha}$. The degrees of freedom required for description of collective motion in $\alpha$-DNS are related to the the motion in relative distance coordinate $R$, to the rotation of system as a whole described by the  angles $\Omega_{R}(\theta_{R},\phi_{R})$, and to the collective rotation of the daughter nucleus described by the  angles $\Omega_{h}(\theta_{h},\phi_{h})$.
The schematic representation of $\alpha$ -DNS with indication of the degrees of freedom used is shown in Fig.~\ref{coor}.

The kinetic energy of the collective motion in DNS is written as  \cite{Shneidman2015}:
  \begin{eqnarray}
    \hat{T}=-\frac{\hbar^2}{2 \mu R^2} \frac{\partial}{\partial R} R^2 \frac{\partial}{\partial R}+
\frac{\hbar^2}{2 \Im_R} \hat{I}_R^2+\frac{\hbar^2}{2 \Im_h(I_{h})}   \hat{I}^2_h,
\label{KinEn-alpha}
  \end{eqnarray}
where angular momentum operators have the form:
\begin{eqnarray}
&& \hat{I}^2_i=-\frac{1}{\sin
\theta_i}\frac{\partial}{\partial\theta_i}\sin \theta_i
\frac{\partial}{\partial\theta_i}-\frac{1}{\sin^2
{\theta_i}}\frac{\partial^2}{\partial \phi_i^2}, \quad(i=R,h).
\label{L-operator}
\end{eqnarray}
In Eq.~\eqref{KinEn-alpha}, $\Im_R=\mu R^2$ is the moments of inertia of rotation of DNS as a whole.
Taking into account that the light fragment of DNS is an $\alpha$-particle, the reduced mass is written as $\mu=m_{0}4(A-4)/A$, where $m_{0}$ is the nucleon mass. In the calculations, it is preferable to use experimental values for energies of yrast states of daughter nucleus. Therefore, the moment of inertia $\Im_h(I_{h})$ is adjusted in such a way that for angular momentum $I_{h}$, the last term of Eq.~\eqref{KinEn-alpha} reproduces the experimental energy of corresponding yrast state of daughter nucleus.

%The angular part of kinetic energy is diagonalized in a set of bipolar spherical functions
%\begin{eqnarray} \label{Basis}
% [Y_{I_{h}}(\Omega_h)\times Y_{I_{R}}(\Omega_R)]_{(I,M)}=
%\sum
%_{M_{h},M_{R}} C_{I_{h}M_{h},I_{R}M_{R}}^{IM}Y_{I_{h}M_{h}}(\Omega_{h})Y_{I_{R}M_{R}}(\Omega_{R}).
%\end{eqnarray}
%Each function in Eq.~\eqref{Basis} describes the state of DNS with total angular momentum $I$ and its projection $M$ with daughter fragment having angular momentum $I_{h}$ and the motion of DNS as a whole with angular momentum $I_{R}$.
Potential energy in general form can be presented as
\begin{eqnarray} \label{eq9}
V(R,{\Omega}_{h},{\Omega}_{R})=\sum_{{\lambda}} V_{{\lambda}}(R)\left[Y_{\lambda}({\Omega}_{h})\times Y_{\lambda}({\Omega}_{R}) \right]_{(00)},
\end{eqnarray}
where bipolar spherical functions are defined as (see Ref.~\cite{Varshalovich1988})
\begin{eqnarray} \label{Basis}
 [Y_{\lambda_{h}}(\Omega_h)\times Y_{\lambda_{R}}(\Omega_R)]_{(\lambda,m)}=
\sum
_{m_{h},m_{R}} C_{\lambda_{h}m_{h},\lambda_{R}m_{R}}^{\lambda m}Y_{\lambda_{h}m_{h}}(\Omega_{h})Y_{\lambda_{R}m_{R}}(\Omega_{R}).
\end{eqnarray}
The functions $\left[Y_{\lambda}({\Omega}_{h})\times Y_{\lambda}({\Omega}_{R}) \right]_{(00)}$ can be related to the Legendre polynomials:
\begin{eqnarray} \label{Angle-conection}
P_{\lambda}(\cos\varepsilon)=\frac{(-1)^{\lambda}}{\sqrt{2{\lambda}+1}}\left[Y_{\lambda}(\Omega_{h})\times Y_{\lambda}(\Omega_{R})\right]_{(00)},
\end{eqnarray}
where $\varepsilon$ is a plane angle between the vector $\bf{R}$ and the symmetry axis of the daughter nucleus (see Fig.~\ref{coor}). Therefore, the angular dependence of potential energy is defined only by angle $\varepsilon$ and expressed as
\begin{eqnarray} \label{Approx-Pot}
V(R,\varepsilon)=\sum_{\lambda} (-1)^{\lambda} \frac{\sqrt{2\lambda+1}}{4\pi} V_{\lambda}(R) P_{\lambda}(\cos\varepsilon).
\end{eqnarray}

\begin{figure}[t]
\begin{center}
\includegraphics[scale=.5 ]{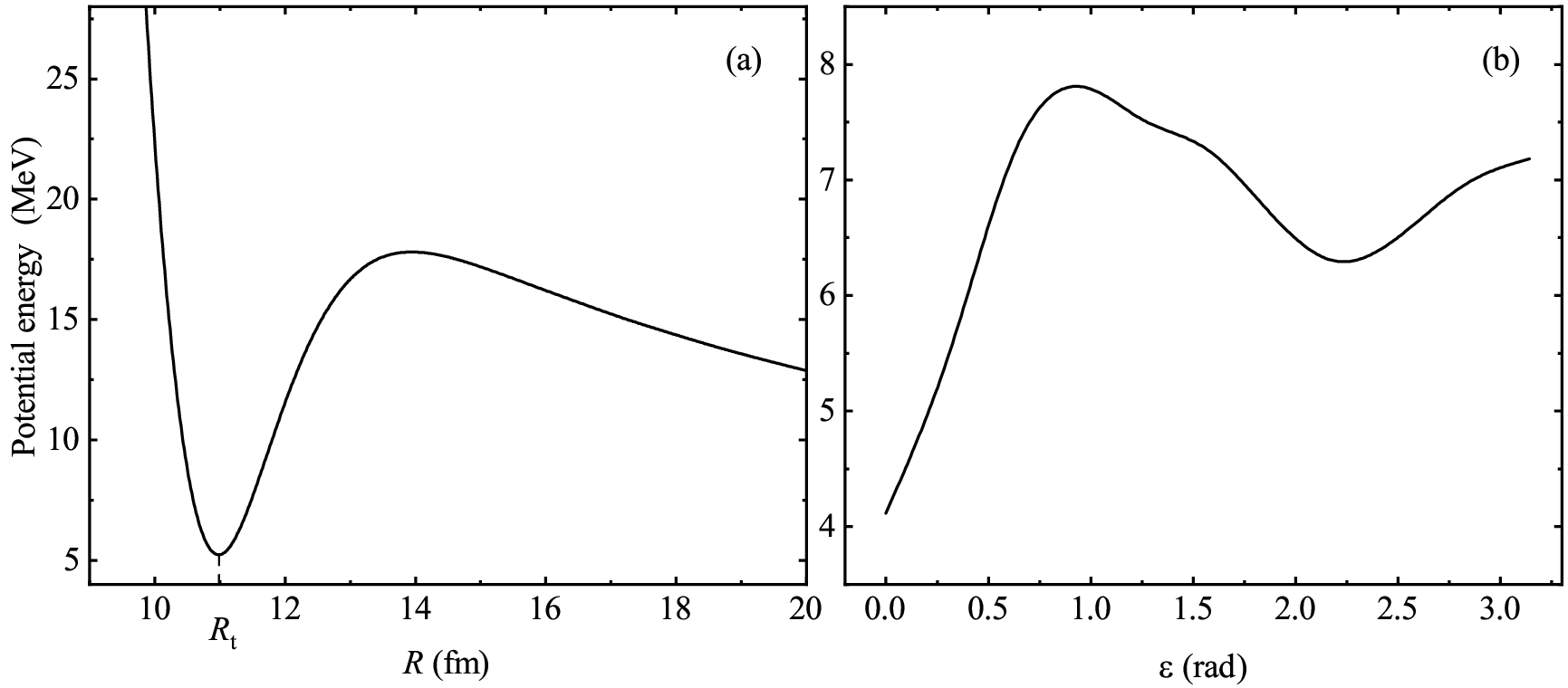}
\end{center}
\caption{Potential energy $V(R,\epsilon)$ of $^{226}$Ra+$\alpha$ DNS. The dependence of potential on relative distance coordinate $R$ for $\epsilon=0$ is shown on panel (a).  The angular dependence of the potential energy in the minimum, $V(R_t(\epsilon),\epsilon)$, is presented in panel (b).}
\label{PotExample}
\end{figure}

An example of potential energy calculation is presented in Fig.~\ref{PotExample}. Panel (a) shows the potential energy as a function of relative distance $R$ calculated for $\varepsilon=0$. As seen, due to interplay between Coulomb and nuclear interactions, the potential energy exhibits a minimum at the distance $R_{t}$ in the vicinity of touching configuration. In Fig.~\ref{PotExample} (b), the potential energy at the minimum of the pocket for each value of $\varepsilon$ is displayed. It is seen that due to nonzero octupole deformation the potential is reflection asymmetric. The potential energy has a minimum at $\varepsilon=0$ with $\alpha$-particle located at the tip of the daughter nucleus. Replacing the potential energy by the oscillator around $R_{t}(\varepsilon=0)$:
\begin{eqnarray} \label{Osc-Pot2}
V(R,\varepsilon)\approx\frac{C_{R}}{2}(R-R_{t}(\varepsilon=0))^2+\frac{C_{\varepsilon}}{2}{\varepsilon}^2,
\end{eqnarray}
we can estimate the frequencies of angular vibration $\hbar\omega_{\varepsilon}$ and vibration in the relative distance coordinate $\hbar\omega_{R}$. Simple estimations show that the motion in $R$ is much faster than the motion in angular variables: $\hbar \omega_R \gg \hbar\omega_{\varepsilon}$.
Hence, for the solution of the Schrodinger equation with Hamiltonian $\hat{H}=\hat{T}+U$, we resort to use of Born-Oppenheimer approximation. In other words, we assume that motion in $R$ is so fast that for each value of angles DNS achieves its equilibrium configuration in $R$ which approximately corresponds to the minimum of pocket at $R_{t}(\varepsilon)$.

The wave function is then factorized as a product of wave functions in slow and fast coordinates:
\begin{eqnarray} \label{B-O-approximation}
\Psi_{\alpha}(R,\Omega_{h},\Omega_{R})=\frac{\psi(R)}{R}\mathcal{Y}(\Omega_{h},\Omega_{R}).
\end{eqnarray}
In the equation for the radial wave function $\psi(R)$, the angular part of the kinetic energy is neglected, which yields the following Schrodinger equation for $\psi(R)$:
\begin{eqnarray}\label{B-O-radial}
\left [-\frac{\hbar^2}{2 \mu } \frac{\partial^2}{\partial R^2}+V(R,\varepsilon) \right ] \psi(R)=E(\varepsilon) \psi(R),
\end{eqnarray}
where $E(\varepsilon)=\langle T_{R}\rangle+ V(R_{t}(\varepsilon),\varepsilon)$.
Here, the average kinetic energy $\langle T_{R}\rangle$ and potential $V$ are calculated at the minimum of the pocket for each value of $\varepsilon$.

Assuming that $\langle T_{R}\rangle$ weakly depends on $\varepsilon$, we write the Hamiltonian for angular vibrations in $\alpha$-DNS in the form:
\begin{eqnarray}\label{Hamiltonian}
\hat{H}&=&\hat{T}+V(\Omega_{h},\Omega_{R}),\nonumber \\
\hat{T}&=&\frac{\hbar^2}{2 \mu R_{t}^2(0)} \hat{I}_R^2+\frac{\hbar^2}{2 \Im_h(I_{h})}   \hat{I}^2_h,\nonumber \\
V(\Omega_{h},\Omega_{R})&=&V(R_{t}(\varepsilon),\varepsilon)=\sum_{{\lambda}} \tilde{V}_{{\lambda}}\left[Y_{\lambda}({\Omega}_{h})\times Y_{\lambda}({\Omega}_{R}) \right]_{(00)},
\end{eqnarray}
where $\tilde{V}_{\lambda}$ represents the  expansion coefficients  of $V(R_{t}(\varepsilon),\varepsilon)$ in terms of bipolar spherical functions.

The Hamiltonian \eqref{Hamiltonian}, is diagonalized on a set of bipolar spherical functions \eqref{Basis}. For the states with $I^{\pi}=0^{+}$ corresponding to the ground state of even-even nuclei the wave functions is simplified as
\begin{eqnarray}   \label{Angle-conection}
\mathcal{Y}(\varepsilon)=\sum_{l}a_{l} \left[Y_{l}(\Omega_{h})\times Y_{l}(\Omega_{R})\right]_{(00)}=\sum_{l}a_{l}(-1)^{l}\sqrt{\frac{2l+1}{2}}P_{l}(\cos\varepsilon).
\end{eqnarray}
The squares of amplitudes $a_l$ give the probabilities to find an $\alpha$-DNS in the state where the relative motion and the daughter nucleus are both excited into the states with angular momentum $l$. Therefore, the dependence of the spectroscopic factor on angular momentum is given by
\begin{eqnarray}
 S_l = S |a_l|^2.
 \end{eqnarray}

\section{Results and discussions}

Using  method described above we perform calculation of angular momentum dependence of spectroscopic factors for even-even actinides.
To perform these calculations the potential energy $V(R,\varepsilon)$ of DNS consisting of a heavy nucleus and an $\alpha$-particle should be specified.
In this work it is calculated as follows:
\begin{eqnarray}
V(R,\epsilon,\beta_2,\beta_3)=U(R,\epsilon,\beta_2,\beta_3)-(B_A-B_{A-4}-B_\alpha),
\label{Pot1}
\end{eqnarray}
where $B_A$, $B_{A-4}$, and $B_\alpha$ are the binding energies of the parent nucleus, daughter nucleus, and $\alpha$-particle, respectively. The ground state deformations and the the experimental values of binding energies are taken from \cite{Moller2016}.
The nucleus-nucleus potential $U(R,\epsilon,\beta_2,\beta_3)$ is taken as the sum of the Coulomb and nuclear interactions $U=U_{C}+U_{N}$. The nuclear part of interaction is calculated using double-folding procedure \cite{Adamian1996} as
\begin{eqnarray}
U_{N}(R,\epsilon,\beta_2,\beta_3)=\int \rho_1({\bf r_1})\rho_1({\bf R_m}-{\bf r_2})F({\bf
r_1}-{\bf r_2})d{\bf r_1} d{\bf r_2}, \label{chap11_poten_E}
\end{eqnarray}
where $F({\bf r_1}-{\bf r_2})$ is the density-dependent effective nucleon–nucleon interaction known as the Migdal forces \cite{Migdal1967}.
The nuclear densities $\rho_i$ are approximated by Fermi distributions with the radius parameter
$r_0$=1.15 fm for heavy fragments and $r_0$=1.0 fm for the $\alpha$-particle. The diffuseness parameter of the $\alpha$-particle density distribution is taken to be 0.48 fm. For heavy clusters, the diffuseness parameters are calculated as $a=0.56 \sqrt{B^{(0)}_n/B_n}$, where $B^{(0)}_n$ and $B_n$ are the neutron binding energies of the nucleus under study and the heaviest isotope of the same element, respectively. Details of the calculations are presented in the works~\cite{Shneidman2015,Shneidman2003}.

\begin{figure}
    \centering
    \includegraphics[width=1.0\linewidth]{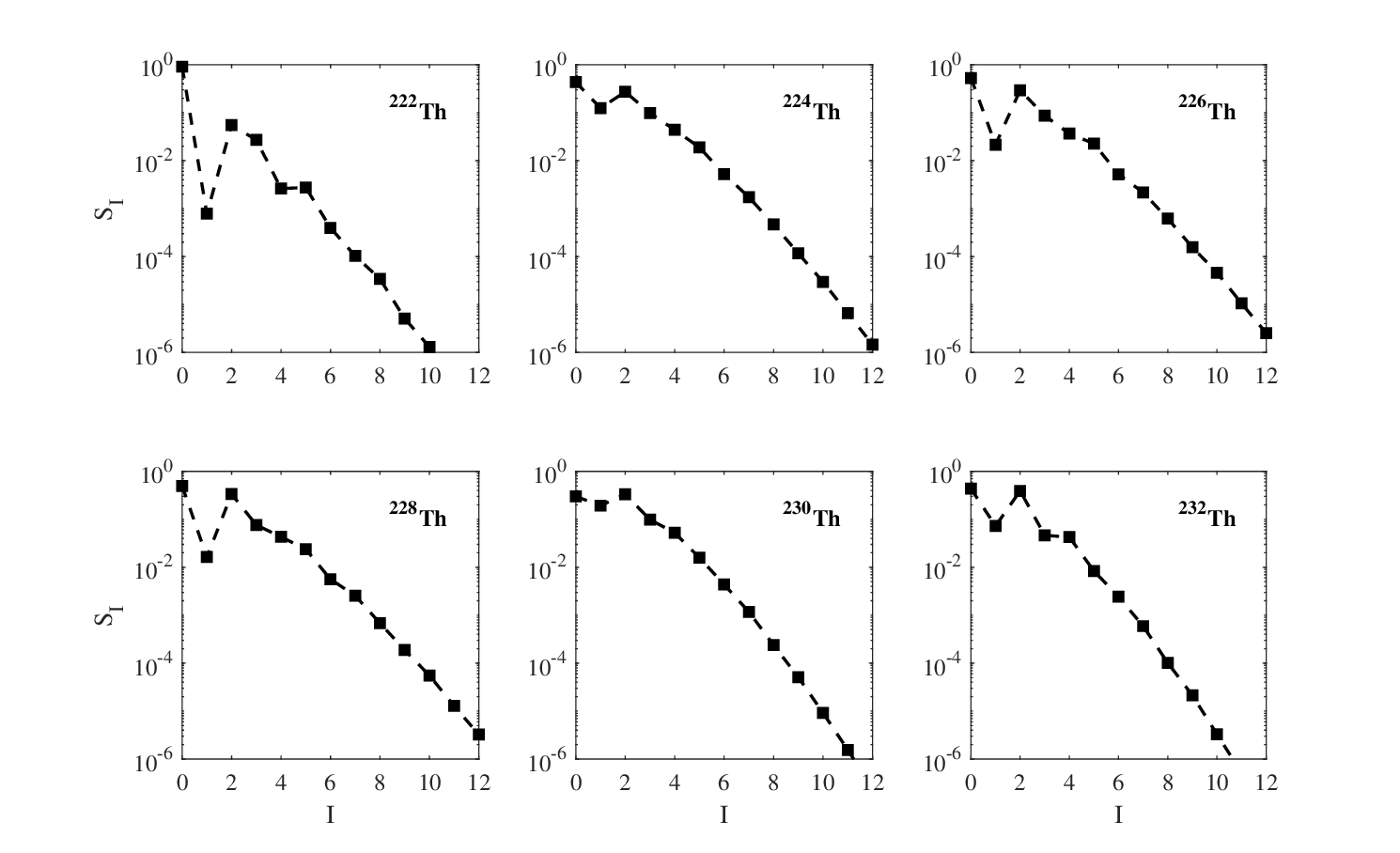}
    \caption{Angular momentum dependence of spectroscopic factors of $\alpha$-decay in various Th isotopes.}
    \label{figTh}
\end{figure}

Results of calculations of  spectroscopic factors $S_I$   are presented in Tables.~\ref{tab-Ra}-\ref{tab-Pu} for $^{224-228}$Ra, $^{222-232}$Th, $^{226-236}$U, and $^{234-242}$Pu.   For Th isotopes the spectroscopic factors are also illustrated in Fig.~\ref{figTh}. To understand the observed behavior of $S_I$, we construct the first order correction to the lowest eigenstate of Hamiltonian \eqref{Hamiltonian}. If potential energy term in \eqref{Hamiltonian} is completely neglected, the ground state of $\hat{H}$ is given by the bipolar function $\left[Y_{0}(\Omega_{h})\times Y_{0}(\Omega_{R})\right]_{(00)}$ only. Therefore, if daughter nucleus is spherical, $\alpha$-particle can only be formed with angular momentum $I=0$, i.e. $S_I=0$ for $I \ne 0$.
The admixture of the other basis states into the ground-state wave function arising due to nonzero deformations can be estimated as
\begin{eqnarray}\label{Perturbation}
a_{I}\approx\frac{\langle \left[Y_{0}(\Omega_{h})\times Y_{0}(\Omega_{R})\right]_{(00)} |V(\Omega_{h},\Omega_{R})| \left[Y_{I}(\Omega_{h})\times Y_{I}(\Omega_{R})\right]_{(00)} \rangle}{E_{d}(I)+\frac{\hbar^2I(I+1)}{2 \mu R_{t}^2(0)}}.
\end{eqnarray}
Here, $E_{d}(I)$ is the energy of excited state of the daughter nucleus.

Since the potential energy depends mainly on small deformation parameters $\beta_{2}$ and $\beta_{3}$, the higher multipoles in the expansion of $V$ in \eqref{Hamiltonian} are strongly suppressed and one expects matrix elements in numerator of Eq.~\eqref{Perturbation} decrease rapidly for $I \geqslant 3$. Moreover, the denominator is increasing with $I$ as $\sim I(I+1)$. These two factors account for the significant reduction of the probability to form an $\alpha$-particle in states with high angular momenta. As illustrated in Fig.~\ref{figTh}, spectroscopic factors decrease substantially with $I$.

One notable observation is that $S_{I}$ exhibits a staggering pattern, with reduced probabilities of forming $\alpha$-particles in states with odd $I$ compared to neighboring states with even $I$. This hindrance is especially pronounced for $I=1$ and gets less significant with increasing angular momentum. The staggering behavior of $S_{I}$ is explained by the effect of parity splitting on $E_{d}(I)$ in denominator of Eq.~\eqref{Perturbation}. Given that in even-even nuclei the states with negative parity and odd angular momenta are shifted up significantly compared to the states with both positive parity and even angular momentum (parity splitting \cite{Jolos2013}), the denominator of Eq.~\eqref{Perturbation} becomes larger for states with odd $I$, generating the staggering trend in $S_I$ for small angular momenta. As angular momentum increases, parity splitting diminishes, leading the even-odd variation in the denominator to vanish, which in turn reduces the staggering effect on $S_I$.

Additional hindrance of the population of $\alpha$-particle in the state with angular momentum $I=1$ is also related to the fact that matrix element in Eq.(\ref{Perturbation}) is significantly reduced in this case. Indeed, the first order contribution to $a_1$ is proportional to $\beta_1 \sim \beta_2 \beta_3$.

It is interesting to compare the observed trend of $S_I$ with the prescription provided by the empirical BD approach. To do so, we assume that the calculated spectroscopic factors are presented as $ S_I = \exp{\left (-c_I E(I)\right )}$, where $E(I)$ is the excitation energy of the yrast state with angular momentum $I$ in the daughter nucleus. The resulting values of $c_I$ are presented in Tables~\ref{tab-Ra}-\ref{tab-Pu}. First, it is evident that $c_{I}$ is not a constant but varies within a large interval between $\sim 5$ MeV$^{-1}$ and $\sim 20$ MeV$^{-1}$ from nucleus to nucleus and for different angular momenta of the same nucleus. This dependence is determined not only by the energies of the excited states of the
daughter nucleus, as is suggested by the Boltzmann distribution, but also by its quadrupole and octupole deformation parameters (see Eq.~\eqref{Perturbation}).

By a closer look to the data presented in Tables~\ref{tab-Ra}-\ref{tab-Pu}, it becomes apparent that the values of $c_{even-I}$ are significantly impacted by quadrupole deformation, whereas $c_{odd-I}$ is influenced by the octupole deformation of the daughter nucleus. Notably, a considerable quadrupole deformation in the daughter nucleus requires the use of BD description with $c_I$ values around (15-20) MeV$^{-1}$. In contrast, for the nuclei with smaller quadrupole deformation, the $c_I \approx (7-8)$ MeV$^{-1}$ are required.
Similarly, if the octupole deformation of daughter nucleus is small,  the $c_{odd-I}$ values are varying in the interval (5-9) MeV$^{-1}$. If, however, the octupole deformation of daughter nucleus is sufficient, like in the case of $^{226,228}$Th, the angular momentum $I=1$ is strongly hindered and $c_I \sim (15-20)$ MeV$^{-1}$.
To summarize, to use BD empirical approach one can take $c_{even-I}\approx c_{odd-I} \approx 8$ MeV$^{-1}$ for nearly spherical nuclei and $c_{even-I}\sim 15$ MeV$^{-1}$, or $c_{odd-I}\sim 16$ MeV$^{-1}$ if, respectively, quadrupole or octupole deformations are significant.\\
{\bf Acknowledgments:}
This research has been funded by the Science Committee of the Ministry of Science and Higher Education of the Republic of Kazakhstan (Grant No. AP19577048).

\begin{table}
  \caption{Calculated spectroscopic factors $S_I$ to form an $\alpha$-particle in a state with angular momentum $I$ and corresponding constants $c_I$ of Boltzmann distribution are presented for different Ra isotopes. The quadrupole $\beta_2$ and octupole $\beta_3$ deformation parameters of  daughter nuclei are taken from \cite{Moller2016}. The energies $E(I)$ of yrast excited states in daughter nuclei are taken from \cite{nndc}.}
    \centering
    \begin{tabular}{|cccc|}
    \hline
         \multicolumn{4}{|c|}{$^{222}$Ra ($\beta_2=0.079, \beta_3=0.139$)}\\ \hline
         $I\quad$&  $E(I)$ (MeV)&  $S_{I}$& $c_{I}$\\ \hline
 0 & 0 & $8.06\times 10^{-1}$ & -\\
 1 & 0.796 & $2.91\times 10^{-3}$ & 7.34 \\
 2 & 0.324 & $9.27\times 10^{-2}$ & 7.34 \\
 3 & 0.84 & $7.36\times 10^{-2}$ & 3.11 \\
 4 & 0.653 & $1.35\times 10^{-2}$ & 6.59 \\
 5 & 1.026 & $8.82\times 10^{-3}$ & 4.61 \\
 6 & 1.014 & $1.91\times 10^{-3}$ & 6.17 \\ \hline
    \end{tabular}
        \begin{tabular}{|cccc|}
    \hline
         \multicolumn{4}{|c|}{$^{224}$Ra ($\beta_2=0.110, \beta_3=0.125$)}\\ \hline
         $I\quad$&  $E(I)$ (MeV)&  $S_{I}$& $c_{I}$\\ \hline
 0 & 0 & $7.27\times 10^{-1}$ & - \\
 1 & 0.645 & $1.09\times 10^{-2}$ & 7.00 \\
 2 & 0.241 & $1.32\times 10^{-1}$ & 8.39 \\
 3 & 0.663 & $9.40\times 10^{-2}$ & 3.57 \\
 4 & 0.534 & $1.97\times 10^{-2}$ & 7.36 \\
 5 & 0.851 & $1.21\times 10^{-2}$ & 5.19 \\
 6 & 0.874 & $2.82\times 10^{-3}$ & 6.72 \\ \hline
    \end{tabular}
        \begin{tabular}{|cccc|}
    \hline
         \multicolumn{4}{|c|}{$^{226}$Ra ($\beta_2=0.110, \beta_3=0.125$)}\\ \hline
         $I\quad$&  $E(I)$ (MeV)&  $S_{I}$& $c_{I}$\\ \hline
 0 & 0 &$ 6.77\times 10^{-1}$ & - \\
 1 & 0.6 & $1.80\times 10^{-2}$ & 6.69
\\
 2 & 0.186 & $1.63\times 10^{-1}$ &           9.75
\\
 3 & 0.636 & $9.89\times 10^{-2}$ &           3.64
\\
 4 & 0.448 & $2.43\times 10^{-2}$ &           8.29
\\
 5 & 0.797 & $1.40\times 10^{-2}$ &           5.36
\\
 6 & 0.768 & $3.38\times 10^{-3}$ &           7.41\\ \hline
    \end{tabular}
        \begin{tabular}{|cccc|}
    \hline
         \multicolumn{4}{|c|}{$^{228}$Ra ($\beta_2=0.142, \beta_3=0.112$)}\\ \hline
         $I\quad$&  $E(I)$ (MeV)&  $S_{I}$& $c_{I}$\\ \hline
0 & 0 & $7.06\times 10^{-1}$ & -\\
 1 & 0.6 & $6.86\times 10^{-3}$ & 8.30 \\
 2 & 0.136 & $2.60\times 10^{-1}$ & 9.91 \\
 3 & 0.651 & $9.73\times 10^{-3}$ & 7.12 \\
 4 & 0.358 & $1.25\times 10^{-2}$ & 12.25 \\
 5 & 0.791 & $4.12\times 10^{-3}$ & 6.94 \\
 6 & 0.641 & $4.53\times 10^{-4}$ & 12.01\\ \hline
    \end{tabular}
          \label{tab-Ra}
\end{table}

\begin{table}
  \caption{Same as in Tab.~\ref{tab-Ra}, but for Th isotopes.}
    \centering
    \begin{tabular}{|cccc|}
    \hline
         \multicolumn{4}{|c|}{$^{222}$Th ($\beta_{2}=0.078, \beta_3=0.125$)}\\ \hline
         $I\quad$&  $E(I)$ (MeV)&  $S_{I}$& $c_{I}$\\ \hline
 0 & 0 & $9.11\times 10^{-1}$ & - \\
 1 & 0.853 &$ 7.83\times 10^{-4}$  & 8.39
\\
 2 & 0.389 & $5.50\times 10^{-2}$ &           7.46
\\
 3 & 0.793 & $2.72\times 10^{-2}$ &           4.55
\\
 4 & 0.741 & $2.62\times 10^{-3}$ &           8.02
\\
 5 & 1.038 & $2.75\times 10^{-3}$ &           5.68
\\
 6 & 1.122 & $3.97\times 10^{-4}$ &           6.98\\ \hline
    \end{tabular}
        \begin{tabular}{|cccc|}
    \hline
         \multicolumn{4}{|c|}{$^{224}$Th ($\beta_{2}=0.111, \beta_3=0.127$)}\\ \hline
         $I\quad$&  $E(I)$ (MeV)&  $S_{I}$& $c_{I}$\\ \hline
0 & 0 & $4.34\times 10^{-1}$ & - \\
 1 & 0.412 & $1.25\times 10^{-1}$ & 5.05
\\
 2 & 0.178 & $2.73\times 10^{-1}$ &           7.30
\\
 3 & 0.474 & $9.82\times 10^{-2}$ &           4.90
\\
 4 & 0.41 & $4.39\times 10^{-2}$ &           7.62
\\
 5 & 0.634 & $1.89\times 10^{-2}$ &           6.26
\\
 6 & 0.688 & $5.22\times 10^{-3}$ &           7.64\\ \hline
    \end{tabular}

        \begin{tabular}{|cccc|}
    \hline
         \multicolumn{4}{|c|}{$^{226}$Th ($\beta_{2}=0.122, \beta_3=0.141$)}\\ \hline
         $I\quad$&  $E(I)$ (MeV)&  $S_{I}$& $c_{I}$\\ \hline
 0 & 0 & $5.30\times 10^{-1}$ & -\\
 1 & 0.242 & $2.14\times 10^{-2}$ & 15.88
\\
 2 & 0.111 & $2.94\times 10^{-1}$ &          11.02
\\
 3 & 0.317 & $8.68\times 10^{-2}$ &           7.71
\\
 4 & 0.301 & $3.67\times 10^{-2}$ &          10.98
\\
 5 & 0.473 & $2.27\times 10^{-2}$ &           8.01
\\
 6 & 0.549 & $5.20\times 10^{-3}$ &           9.58\\ \hline
    \end{tabular}
        \begin{tabular}{|cccc|}
    \hline
         \multicolumn{4}{|c|}{$^{228}$Th ($\beta_{2}=0.143, \beta_3=0.139$)}\\ \hline
         $I\quad$&  $E(I)$ (MeV)&  $S_{I}$& $c_{I}$\\ \hline
0 & 0 & $4.96\times 10^{-1}$ & -\\
 1 & 0.215 & $1.63\times 10^{-2}$ & 19.14
\\
 2 & 0.084 & $3.36\times 10^{-1}$ &          12.98
\\
 3 & 0.29 & $7.60\times 10^{-2}$ &           8.89
\\
 4 & 0.25 & $4.31\times 10^{-2}$ &          12.58
\\
 5 & 0.433 & $2.37\times 10^{-2}$ &           8.64
\\
 6 & 0.479 & $5.63\times 10^{-3}$ &          10.81\\ \hline
    \end{tabular}

        \begin{tabular}{|cccc|}
    \hline
         \multicolumn{4}{|c|}{$^{230}$Th ($\beta_{2}=0.164, \beta_3=0.112$)}\\ \hline
         $I\quad$&  $E(I)$ (MeV)&  $S_{I}$& $c_{I}$\\ \hline
 0 & 0 & $3.01\times 10^{-1}$ & - \\
 1 & 0.253 & $1.93\times 10^{-1}$ &  6.51
\\
 2 & 0.067 & $3.34\times 10^{-1}$ &          16.35
\\
 3 & 0.321 & $9.82\times 10^{-2}$ &           7.23
\\
 4 & 0.211 & $5.25\times 10^{-2} $&          13.96
\\
 5 & 0.446 & $1.58\times 10^{-2}$ &           9.30
\\
 6 & 0.416 & $4.36\times 10^{-3}$ &          13.06\\ \hline
    \end{tabular}
        \begin{tabular}{|cccc|}
    \hline
         \multicolumn{4}{|c|}{$^{232}$Th ($\beta_{2}=0.174, \beta_3=0.083$)}\\ \hline
         $I\quad$&  $E(I)$ (MeV)&  $S_{I}$& $c_{I}$\\ \hline
 0 & 0 & $4.35\times 10^{-1}$ & - \\
 1 & 0.474 & $7.25\times 10^{-2}$ &  5.54
\\
 2 & 0.063 & $3.92\times 10^{-1}$ &          14.87
\\
 3 & 0.538 & $4.63\times 10^{-2}$ &           5.71
\\
 4 & 0.205 & $4.26\times 10^{-2}$ &          15.40
\\
 5 & 0.656 & $8.35\times 10^{-3}$ &           7.29
\\
 6 & 0.412 & $2.43\times 10^{-3}$ &          14.61\\\hline
    \end{tabular}
    \label{tab-Th}
\end{table}

\begin{table}
  \caption{Same as in Tab.~\ref{tab-Ra}, but for U isotopes.}
    \centering
    \begin{tabular}{|cccc|}
    \hline
         \multicolumn{4}{|c|}{$^{226}$U ($\beta_2=0.111, \beta_3=0.140$)}\\ \hline
         $I\quad$&  $E(I)$ (MeV)&  $S_{I}$& $c_{I}$\\ \hline
 0 & 0 & $6.18\times 10^{-1}$ & - \\
 1 & 0.246 & $1.78\times 10^{-2}$ & 16.39 \\
 2 & 0.183 & $2.39\times 10^{-1}$ & 7.83 \\
 3 & 0.467 & $7.63\times 10^{-2}$ & 5.51 \\
 4 & 0.439 & $2.65\times 10^{-2}$ & 8.27 \\
 5 & 0.65 & $1.71\times 10^{-2}$ & 6.26 \\
 6 & 0.749 & $3.68\times 10^{-3}$ & 7.48 \\ \hline
    \end{tabular}
        \begin{tabular}{|cccc|}
    \hline
         \multicolumn{4}{|c|}{$^{228}$U ($\beta_2=0.144, \beta_3=0.153$)}\\ \hline
         $I\quad$&  $E(I)$ (MeV)&  $S_{I}$& $c_{I}$\\ \hline
 0 & 0 & $3.58\times 10^{-1}$ & - \\
 1 & 0.251 & $6.22\times 10^{-2}$ & 11.06 \\
 2 & 0.098 & $3.43\times 10^{-1}$ & 10.92 \\
 3 & 0.305 & $1.17\times 10^{-1}$ & 7.05 \\
 4 & 0.284 & $6.60\times 10^{-2}$ & 9.57 \\
 5 & 0.465 & $3.61\times 10^{-2}$ & 7.14 \\
 6 & 0.535 & $1.09\times 10^{-2}$ & 8.45 \\\hline
    \end{tabular}
        \begin{tabular}{|cccc|}
    \hline
         \multicolumn{4}{|c|}{$^{230}$U ($\beta_2=0.154, \beta_3=0.139$)}\\ \hline
         $I\quad$&  $E(I)$ (MeV)&  $S_{I}$& $c_{I}$\\ \hline
 0 & 0 &$ 3.55\times 10^{-1}$ & -\\
 1 & 0.23 & $5.42\times 10^{-2}$ & 12.68 \\
 2 & 0.072 & $3.73\times 10^{-1}$ & 13.68 \\
 3 & 0.308 & $9.86\times 10^{-2}$ & 7.52 \\
 4 & 0.226 & $6.98\times 10^{-2}$ & 11.78 \\
 5 & 0.451 & $3.29\times 10^{-2}$ & 7.57 \\
 6 & 0.447 & $1.03\times 10^{-2}$ & 10.24 \\ \hline
    \end{tabular}
        \begin{tabular}{|cccc|}
    \hline
         \multicolumn{4}{|c|}{$^{232}$U ($\beta_2=0.174, \beta_3=0.111$)}\\ \hline
         $I\quad$&  $E(I)$ (MeV)&  $S_{I}$& $c_{I}$\\ \hline
0 & 0 & $4.09\times 10^{-1}$ & - \\
 1 & 0.328 & $1.34\times 10^{-2}$ & 13.15 \\
 2 & 0.058 & $4.28\times 10^{-1}$ & 14.63 \\
 3 & 0.396 &$ 5.10\times 10^{-2}$ & 7.52 \\
 4 & 0.187 & $6.52\times 10^{-2}$ & 14.60 \\
 5 & 0.519 & $2.17\times 10^{-2}$ & 7.38 \\
 6 & 0.378 & $7.17\times 10^{-3}$ & 13.06\\ \hline
    \end{tabular}
     \begin{tabular}{|cccc|}
    \hline
         \multicolumn{4}{|c|}{$^{234}$U ($\beta_2=0.95, \beta_3=0.050$)}\\ \hline
         $I\quad$&  $E(I)$ (MeV)&  $S_{I}$& $c_{I}$\\ \hline
 0 & 0 & $4.42\times 10^{-1}$ & - \\
 1 & 0.508 & $3.68\times 10^{-5}$ & 20.10 \\
 2 & 0.053 &$ 4.82\times 10^{-1}$ & 13.78 \\
 3 & 0.572 & $5.29\times 10^{-3}$ & 9.17 \\
 4 & 0.174 & $6.25\times 10^{-2}$ & 15.94 \\
 5 & 0.687 & $3.56\times 10^{-3}$ & 8.21 \\
 6 & 0.357 & $3.88\times 10^{-3}$ & 15.55\\  \hline
    \end{tabular}
        \begin{tabular}{|cccc|}
    \hline
         \multicolumn{4}{|c|}{$^{236}$U ($\beta_2=0.205, \beta_3=0.050$)}\\ \hline
         $I\quad$&  $E(I)$ (MeV)&  $S_{I}$& $c_{I}$\\ \hline
0 & 0 & $4.22\times 10^{-1}$ & - \\
 1 & 0.714 & $4.07\times 10^{-6}$ & 17.38 \\
 2 & 0.049 & $4.96\times 10^{-1}$ & 14.33 \\
 3 & 0.774 & $3.51\times 10^{-3}$ & 7.30 \\
 4 & 0.162 & $7.01\times 10^{-2}$ & 16.41 \\
 5 & 0.884 & $2.96\times 10^{-3}$ & 6.59 \\
 6 & 0.333 & $4.66\times 10^{-3}$ & 16.12 \\\hline
    \end{tabular}
          \label{tab-U}
\end{table}

\begin{table}
  \caption{Same as in Tab.~\ref{tab-Ra} but for Pu isotopes.}
    \centering
    \begin{tabular}{|cccc|}
    \hline
         \multicolumn{4}{|c|}{$^{234}$Pu ($\beta_2=0.185, \beta_3=0.050$)}\\ \hline
         $I\quad$&  $E(I)$ (MeV)&  $S_{I}$& $c_{I}$\\ \hline
 0 & 0 & $5.13\times 10^{-1}$ & -\\
 1 & 0.367 & $1.52\times 10^{-4}$ & 23.95 \\
 2 & 0.051 & $4.43\times 10^{-1}$ & 15.97 \\
 3 & 0.435 &$ 4.00\times 10^{-3}$ & 12.70 \\
 4 & 0.169 & $3.62\times 10^{-2}$ & 19.65 \\
 5 & 0.558 & $2.07\times 10^{-3}$ & 11.8 \\
 6 & 0.347 & $1.24\times 10^{-3}$ & 19.28 \\ \hline
    \end{tabular}
        \begin{tabular}{|cccc|}
    \hline
         \multicolumn{4}{|c|}{$^{236}$Pu ($\beta_2=0.206, \beta_3=0.050$)}\\ \hline
         $I\quad$&  $E(I)$ (MeV)&  $S_{I}$& $c_{I}$\\ \hline
 0 & 0 & $4.69\times 10^{-1}$ & -\\
 1 & 0.786 & $7.50\times 10^{-4}$ & 9.15 \\
 2 & 0.048 & $4.73\times 10^{-1}$ & 15.58 \\
 3 & 0.849 & $3.31\times 10^{-3}$ & 6.73 \\
 4 & 0.157 & $4.79\times 10^{-2}$ & 19.35 \\
 5 & 0.962 & $2.88\times 10^{-3}$ & 6.08 \\
 6 & 0.323 & $2.09\times 10^{-3}$ & 19.10 \\ \hline
    \end{tabular}
        \begin{tabular}{|cccc|}
    \hline
         \multicolumn{4}{|c|}{$^{238}$Pu ($\beta_2=0.215, \beta_3=0.050$)}\\ \hline
         $I\quad$&  $E(I)$ (MeV)&  $S_{I}$& $c_{I}$\\ \hline
 0 & 0 & $4.48\times 10^{-1}$ & - \\
 1 & 0.786 & $6.23\times 10^{-4}$ & 9.39 \\
 2 & 0.043 & $4.87\times 10^{-1}$ & 16.72 \\
 3 & 0.849 & $3.45\times 10^{-3}$ & 6.68 \\
 4 & 0.143 & $5.45\times 10^{-2}$ & 20.34 \\
 5 & 0.962 & $3.20\times 10^{-3}$ & 5.97\\
 6 & 0.296 & $2.61717\times 10^{-3}$ & 20.0867 \\\hline
    \end{tabular}
        \begin{tabular}{|cccc|}
    \hline
         \multicolumn{4}{|c|}{$^{240}$Pu ($\beta_2=0.226, \beta_3=0.050$)}\\ \hline
         $I\quad$&  $E(I)$ (MeV)&  $S_{I}$& $c_{I}$\\ \hline
0 & 0 & $4.35\times 10^{-1}$ & -\\
 1 & 0.688 & $3.56\times 10^{-4}$ & 11.54 \\
 2 & 0.045 & $4.98\times 10^{-1}$ & 15.48 \\
 3 & 0.744 & $2.00\times 10^{-3}$ & 8.35 \\
 4 & 0.149 & $5.89\times 10^{-2}$ & 19.00 \\
 5 & 0.848 & $1.92\times 10^{-3}$ & 7.38 \\
 6 & 0.31 & $2.81\times 10^{-3}$ & 18.95 \\ \hline
    \end{tabular}
     \begin{tabular}{|cccc|}
    \hline
         \multicolumn{4}{|c|}{$^{242}$Pu ($\beta_2=0.236, \beta_3=0.050$)}\\ \hline
         $I\quad$&  $E(I)$ (MeV)&  $S_{I}$& $c_{I}$\\ \hline
 0 & 0 & $4.15\times 10^{-1}$ & - \\
 1 & 0.68 & $3.09\times 10^{-4}$ & 11.88 \\
 2 & 0.045 & $5.09\times 10^{-1}$ & 14.99 \\
 3 & 0.732 & $2.01\times 10^{-3}$ & 8.48 \\
 4 & 0.148 & $6.70\times 10^{-2}$ & 18.26 \\
 5 & 0.827 & $2.07\times 10^{-3}$ & 7.47 \\
 6 & 0.307 & $3.58\times 10^{-3}$ & 18.35 \\\hline
    \end{tabular}

          \label{tab-Pu}
\end{table}

\end{document}